\documentclass[conference]{IEEEtran}
\usepackage{cite}

\usepackage[cmex10]{amsmath}

\usepackage{array}

\usepackage{xfrac}
\usepackage{amsfonts}
\usepackage{amssymb}
\usepackage{amsbsy}
\usepackage{amsthm}
\usepackage{longtable}
\usepackage{cite}

 \usepackage{graphicx}
\usepackage{tabularx}

\newtheorem{theorem}{Theorem}

\begin{document}
\title{Joint Source-Channel Coding for Broadcast Channel with Cooperating Receivers }


\author{\IEEEauthorblockN{Sajjad~Bahrami\IEEEauthorrefmark{1},
Behrooz~Razeghi\IEEEauthorrefmark{1},
Mostafa~Monemizadeh\IEEEauthorrefmark{2} and
Ghosheh~Abed~Hodtani\IEEEauthorrefmark{1}
}
\IEEEauthorblockA{\IEEEauthorrefmark{1}Department of Electrical Engineering, Ferdowsi University of Mashhad, Mashhad, Iran}
\IEEEauthorblockA{\IEEEauthorrefmark{2}Department of Electrical Engineering, University of Neyshabur, Neyshabur, Iran}
\IEEEauthorblockA{
Email: sajjad.bahrami.ir@ieee.org, behrooz.razeghi.r@ieee.org, mostafamonemizadeh@gmail.com, hodtani@um.ac.ir}
}

\maketitle

\begin{abstract} 
It is known that, as opposed to point-to-point channel, separate source and channel coding is not optimal in general for sending correlated sources over multiuser channels. In some works joint source-channel coding has been investigated for some certain multiuser channels, e.g., multiple access channel (MAC) and broadcast channel (BC). In this paper, we obtain a sufficient condition for transmitting arbitrarily correlated sources over a discrete memoryless BC with cooperating receivers, where the receivers are allowed to exchange messages via a pair of noisy cooperative links. It is seen that our result is a general form of previous ones and includes them as its special cases.  
\end{abstract}

%

\IEEEpeerreviewmaketitle

\section{Introduction}
Cover, El Gamal and Salehi first showed that separate source and channel coding is not optimal in general for transmitting correlated sources over multiuser channels because of not utilizing the correlation between sources in generating channel input codewords \cite{Salehi:MAC}. Hence, they used a correlation preserving codebook generation technique and obtained a sufficient condition for a discrete memoryless multiple access channel (DM-MAC) with arbitrarily correlated sources which includes its previously known separate source and channel coding results for the DM-MAC as special cases. Later, Han and Costa \cite{Han:BC} obtained a sufficient condition (amended by Kramer and Nair \cite{Kramer:BC}) for the problem of sending arbitrarily correlated sources over a DM broadcast channel (DM-BC). The joint source--channel coding proposed by Han and Costa includes Marton's coding for the DM-BC with independent messages \cite{Marton:BC} as a special case. More recently, the joint--source channel coding problem has been studied for transmitting arbitrarily correlated sources over some other multiuser channels and necessary and sufficient conditions have been provided for them (\cite{Gündüz}-\cite{MARC}). It is worth noting that, in general, for multiuser channels with distributed transmitters such as the MAC, joint source--channel coding causes a cooperation between the transmitters through generating correlation--preserving codebook and thereby, helps communication. For multiuser channels with one transmitter (or centeralized transmitters) such as the BC, joint source--channel coding makes use of the compatibility between the sources and channel and thereby, helps communication.  

It is known that cooperation between nodes, as an efficient approach to increase the rate without any extra spectrum allocation cost, can help communication. One of the related works that has considered cooperation between nodes is \cite{Servetto}, in which the authors studied the BC with \textit{independent messages} and cooperating receivers that cooperate over two \textit{noiseless} conferencing links (the case of noiseless relay). They found the capacity region for the case of the physically degraded BC as well as presented some achievable regions for the case of the general BC by using the results presented in \cite{Relay} and \cite{Meulen}. Another relevant work is \cite{RBC}, where the authors derived capacity bounds for the discrete memoryless partially cooperative relay broadcast channel (PCRBC) with independent messages and established the capacity regions for some special classes of it such as the semi-deterministic PCRBC, the orthogonal PCRBC, and the parallel relay channel with unmatched degraded subchannels.

\begin{figure}
\label{Fig1}
\centering
\includegraphics[scale=0.56]{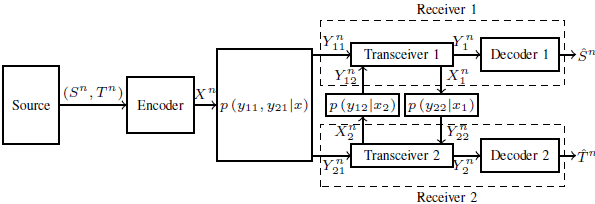}
\caption{Transmission of two arbitrarily correlated sources over a broadcast channel with noisy cooperative links between receivers.}
\end{figure}

In this paper, two important topics in network information theory are investigated: (i) correlation between sources and (ii) cooperation between receivers. Correlation between sources may result from the correlation between observations of different users. On the other hand, cooperation between receivers is due to the inherent broadcasting nature of wireless networks with interactive nodes. Both of these phenomena are seen in wireless sensor and ad-hoc networks. We here consider a communication channel including these two issues. Specifically, we study the problem of sending arbitrarily correlated sources over a general DM-BC with two cooperating receivers that cooperate over a pair of discrete memoryless noisy channels (Fig. \ref{Fig1}). For such channel, we propose a joint source--channel coding and thereby, we obtain a sufficient condition for transmitting arbitrarily correlated sources over the channel. 

The rest of the paper is organized as follows. In Section II we present preliminaries and definitions. Section III is devoted to the main results. The proof of Theorem 1 is given in Section IV. In Section V we conclude the paper.

\section{Preliminaries and Definitions }
In this section we present some basic concepts which will be used in the following sections of this paper. First, we describe the notation as follows.
\vspace{+.15cm}

\textit{Notation:} Throughout the paper, random variables and their realizations are denoted by uppercase and lowercase letters, respectively. $T_{\delta}^{(n)}(X) $ represents the \textit{strongly typical} set of $n$-sequences $x^n$. Random variables $S\in \mathcal{S}$ and $T\in \mathcal{T}$ are used to denote two arbitrarily correlated sources with joint distribution $p\left( s,t\right)$. Alphabets $\mathcal{S}$ and $\mathcal{T}$ are assumed to be finite sets. In addition, the sources $S$ and $T$ are assumed to be memoryless, i.e.,
 {\small \begin{equation}
 p\left( s^n,t^n\right) =\prod _{m=1}^np\left( s_m,t_m\right) .\nonumber
 \end{equation}
 }As shown in Fig. \ref{Fig1}, random variables $X$, $Y_{11}$ and $Y_{21}$, $X_1$ and $X_2$, $Y_{12}$ and $Y_{22}$ denote channel input, channel outputs, inputs of cooperative links, outputs of cooperative links, respectively. Also, $Y_1=\left( Y_{11},Y_{12}\right)$ and $Y_2=\left( Y_{21},Y_{22}\right)$ represent inputs of decoders at the receivers.  

We now introduce some needed definitions.

\vspace{+.1cm}
\textit{Definition 1:} A 2-receiver discrete memoryless broadcast channel with two noisy cooperative links between receivers (shown in Fig. \ref{Fig1}) consists of a channel input alphabet $\mathcal{X}$, two cooperative-link input alphabets $\mathcal{X} _1, \mathcal{X} _2$, two cooperative-link output alphabets $\mathcal{Y} _{12}, \mathcal{Y} _{22}$, two channel output alphabets $\mathcal{Y}_{11}, \mathcal{Y}_{21}$, and conditional probability mass functions $p\left( y_{11},y_{21}|x\right)$, $p\left( y_{22}|x_1\right)$, $p\left( y_{12}|x_2\right)$. The channel and cooperative links are memoryless, i.e.,
{\small \begin{align}
p\left( y_{11}^n,y_{21}^n|x^n\right) &= \prod _{m=1}^np\left( y_{{11}_m},y_{{21}_m}|x_m\right) ,\nonumber\\
p\left( y_{22}^n|x^n_1\right) &= \prod _{m=1}^np\left( y_{{22}_m}|x_{{1_m}}\right) ,\nonumber\\
p\left( y_{12}^n|x^n_2\right) &= \prod _{m=1}^np\left( y_{{12}_m}|x_{{2_m}}\right) .\nonumber
\end{align}
}All alphabets $\mathcal{X}, \mathcal{Y} _{11}, \mathcal{Y} _{21}, \mathcal{X} _1, \mathcal{X} _2, \mathcal{Y} _{12}$ and $\mathcal{Y} _{22}$ are finite sets. 
Here, the transmitter wants to send the source sequence $S^n$ to receiver 1 and the source sequence $T^n$ to receiver 2. In the DM-BC with cooperating receivers shown in Fig. \ref{Fig1}, each receiver acts as a relay to help the other one recover its message correctly. Here, we assume a \textit{one--round} cooperation scheme like a single-step conference that has less delay in comparison to a two-step conference \cite{Servetto}. In the single-step conference, as opposed to the two-step case, both receivers send their cooperative messages simultaneously and do not use their decoded information in order to encode conference messages.

\vspace{+.15cm}
\textit{Definition 2:} An $\left(n,\lambda \right) $ joint source-channel code for the channel consists of an encoding function
{\small \begin{equation}
 e: \mathcal{S}^n\times \mathcal{T}^n\longrightarrow \mathcal{X}^n, \nonumber
\end{equation}
}which maps each source output pair $\left( s^n,t^n\right) $ to a channel input $x^n$ as $x^n=e\left( s^n,t^n\right) $, two sets of relay functions $\{f_{{1_m}}\}_{m=1}^n$ and $\{f_{{2_m}}\}_{m=1}^n$ such that for $1\leq m \leq n$
{\small \begin{align}
x_{{1_m}}&=f_{{1_m}} \left( y_{11}(1),y_{11}(2),\cdots,y_{11}(m-1)\right), \nonumber\\
x_{{2_m}}&=f_{{2_m}} \left( y_{21}(1),y_{21}(2),\cdots,y_{21}(m-1)\right),
\nonumber
\end{align}
}and two decoding functions as 
{\small \begin{align}
g_1&: \mathcal{Y}_{11}^n \times \mathcal{Y}_{12}^n\longrightarrow \mathcal{S}^n,  \nonumber\\
g_2&: \mathcal{Y}_{21}^n \times \mathcal{Y}_{22}^n\longrightarrow \mathcal{T}^n,
\nonumber
\end{align}
}such that
{\small \begin{equation}
P_e^{(n)}=Pr\{ \hat{s}^n \neq s^n ~~ \mathrm{or} ~~ \hat{t}^n \neq t^n\} \leq \lambda ,\nonumber
\end{equation}
}where, $P_e^{(n)}$ is the average probability of error and $\hat{s}^n=g_1\left( y_{11}^n , y_{12}^n\right) $ and $\hat{t}^n=g_2\left( y_{21}^n , y_{22}^n\right)$ are recovered information at the receivers 1 and 2, respectively.

\vspace{+.15cm}
\textit{Definition 3:} A pair of sources $\left( S,T\right) $ is called admissible for the DM-BC with cooperating receivers (shown in Fig. \ref{Fig1}), if for any given $0<\lambda <1$ and for any sufficiently large $n$, there exists an $\left( n,\lambda \right) $ joint source-channel code. The set of all admissible sources $\left( S,T\right) $ is called admissible source region for this channel.    

As stated in \cite{Han:BC}, finding necessary and sufficient condition for a pair of sources $\left( S,T\right) $ to belong to admissible source region of a DM-BC is difficult, therefore in this paper, we only try to find a sufficient condition for the correlated sources $\left( S,T\right) $ to belong to admissible source region of the DM-BC with cooperating receivers. 

\section{Main Results}
In this section, we derive a sufficient condition for the arbitrarily correlated sources $\left( S,T\right) $ to be admissible for the DM-BC with cooperating receivers defined above. Let $K=\alpha \left( S\right) =\beta \left( T\right) $ denote the common part of the source variables $S$ and $T$ in the sense of G\'{a}cs and K\"{o}rner \cite{ElGamal:book}. Here, we consider auxiliary random variables $W$, $U$ and $V$ with finite alphabets $\mathcal{W}$, $\mathcal{U}$ and $\mathcal{V}$, respectively, such that the following Markov chain holds.
\begin{equation}
\label{Markov}
\left(S,T\right) \longrightarrow \left(W,U,V\right)\longrightarrow X\longrightarrow \left(Y_{11},Y_{12},Y_{21},Y_{22}\right) .
\end{equation}
Note that the auxiliary variable $W$ carries the common information of $S$ and $T$, and auxiliary variables $U$ and $V$ carry the private information of $S$ and $T$, respectively. 

\vspace{+.25cm}

\begin{theorem}
A source pair $(S,T)$ is admissible for the 2-receiver DM-BC with cooperating receivers (shown in Fig. \ref{Fig1} and defined above) if there exists auxiliary random variables $W$, $U$ and $V$ that satisfy the Markov chain \eqref{Markov} and the inequalities
{\small \begin{align}
H(S) &< I(SWU;\hat{Y}_2Y_{11}Y_{12}) -I(T;WU|S),  \\
H(T) &< I(TWV;\hat{Y}_1Y_{21}Y_{22}) -I(S;WV|T), \\
H(S,T) &< \min \Big\{I(KW;\hat{Y}_2Y_{11}Y_{12}) ,I(KW;\hat{Y}_1Y_{21}Y_{22}) \Big\} \nonumber \\
&~~~+I(SU;\hat{Y}_2Y_{11}Y_{12}|KW) +I(TV;\hat{Y}_1Y_{21}Y_{22}|KW) \nonumber \\
&~~~-I(SU;TV|KW), \\
H(S,T) &< I(SWU;\hat{Y}_2Y_{11}Y_{12}) +I(TWV;\hat{Y}_1Y_{21}Y_{22}) \nonumber \\
 &~~~-I(SU;TV|KW) -I(ST;KW) , 
\end{align} 
}subject to
{\small \begin{align}
I( X_1;Y_{22}) &\geq I(\hat{Y}_1;Y_{11}|X_1) - I(\hat{Y}_1;Y_{21}Y_{22}|X_1),~~~~~~   \\
I( X_2;Y_{12}) &\geq I(\hat{Y}_2;Y_{21}|X_2) - I(\hat{Y}_2;Y_{11}Y_{12}|X_2). 
\label{theorem1}
\end{align}
}Note that the joint distribution of the channel and auxiliary random variables is as
{\small \begin{IEEEeqnarray}{l}
\label{distribution}
p(s,t,w,u,v,x,x_1,x_2,y_{11},y_{21},y_{12},y_{22},\hat{y}_1,\hat{y}_2) =p(s,t) \nonumber \\
~~~\times p(w,u,v|s,t) p(x|w,u,v) p(x_1) p(x_2) p(y_{11},y_{21}|x)   \nonumber \\
~~~~~~\times p(y_{12}|x_2) p(y_{22}|x_1) p(\hat{y}_1|y_{11},x_1) p(\hat{y}_2|y_{21},x_2) . 
\end{IEEEeqnarray}
}In the above relations, $\hat{Y}_1$ (an estimate of $Y_{11}$) and $\hat{Y}_2$ (an estimate of $Y_{21}$) are random variables to help receivers in final step of detection and their role will become more clear in the proof of Theorem.
\end{theorem}

\textit{Proof:} Refer to Section IV.

\vspace{+.15cm}

Now, we mention some special cases of the obtained admissible region to demonstrate the breadth of our main result. 

\vspace{+.10cm}
\begin{enumerate}
\item \textit{Non-cooperative BC with arbitrarily correlated sources:} By removing cooperative links, letting $X_1=X_2=Y_{12}=Y_{22}=\hat{Y}_1=\hat{Y}_2=\phi $, $Y_1=Y_{11}, Y_2=Y_{21}$, and considering the distribution \eqref{distribution.Han} instead of the distribution \eqref{distribution}, the sufficient condition presented in Theorem 1 boils down to that for the non-cooperative BC with arbitrarily correlated sources (inequalities (2)-(5) in \cite{Kramer:BC}).
{\small 
\begin{align}
\label{distribution.Han}
p(s,t,w,u,v,x,y_{1},y_{2})&=p(s,t)p(w,u,v|s,t) ~~~~~\nonumber \\
  &~~~\times p(x|w,u,v) p(y_{1},y_{2}|x) . 
\end{align}}
\item \textit{Cooperative BC with noiseless conferencing links and independent sources:} Let $S$ and $T$ be two independent sources and set $H\left( S\right) =R_1$ and $H\left( T\right) =R_2$. Note that in this case, $H\left( S,T\right) =H\left( S\right) +H\left( T\right) =R_1+R_2$ and $K=W=\phi$. Make cooperative links noiseless with the capacities $C_{12}$ and $C_{21}$. Moreover, let $X_1=Y_{22}=\phi$, $X_2=Y_{12}=\phi$, $Y_1=Y_{11}, Y_2=Y_{21}$ and consider the distribution \eqref{distribution.Servetto} instead of the distribution \eqref{distribution}. Under such conditions, the sufficient condition in Theorem 1 reduces to the achievable rate region for the cooperative BC with independent sources, where the receivers cooperate over two noiseless conferencing links (the result presented in Theorem 2 in \cite{Servetto}).
{\small 
\begin{IEEEeqnarray}{l}
\label{distribution.Servetto}
p(u,v,x,y_{1},y_{2},\hat{y}_1,\hat{y}_2)=p(u,v) p(x|u,v) ~~~~~~~~~~~~~~\nonumber \\
 ~~~~~~~~~~~~~~~~~~~~~~~ \times  p(y_{1},y_{2}|x) p(\hat{y}_1|y_{1}) p(\hat{y}_2|y_{2}). 
\end{IEEEeqnarray}}
\end{enumerate}

\section{Proof of Theorem 1}
\subsection{Encoding scheme}
We here present a joint source-channel coding scheme which is based on estimate-and-forward (EAF) strategy and uses block Markov encoding and random partitioning. First, let us take a brief look at joint source-channel coding at the transmitter, then we explain the coding at the receivers. 
\vspace{+.10cm}

\textit{Random partitioning:} For each source output $s^n\in \mathcal{S}^n$ and $t^n\in \mathcal{T}^n$ we assign a random index $\theta=\sigma \left( s^n\right)$ and $\varphi=\tau \left( t^n\right) $, respectively. These indices are with equal probability $2^{-nr_1}$ and $2^{-nr_2}$, where $2^{nr_1}$ and $2^{nr_2}$ denote the number of indices $\theta $ and $\varphi $, respectively. \vspace{+.10cm}

\textit{Generation of random codes:} Suppose that a joint distribution factorized as \eqref{distribution} and satisfying conditions \eqref{Markov}-\eqref{theorem1} in Theorem 1 is given. For each $\theta $, $\varphi $ and $k^n=\alpha \left( s^n\right)=\beta \left( t^n\right)$, where $K$ is common variable between $S$ and $T$ with alphabet $\mathcal{K}$, we generate $2^{n\rho _0}$ independent $n$-sequences $w^n_{\theta ,\varphi}\left( k^n\right) \in \mathcal{W}^n$ according to $\prod _{m=1}^np\left( w_{{\theta , \varphi} _m}|k_m\right) $. Next for each pair {\small $\left( s^n,w^n\right) $} we generate $2^{n\rho_1}$ independent $n$-sequences $u^n\left( s^n,w^n\right) \in \mathcal{U}^n$ according to {\small $\prod_{m=1}^n p\left( u_m |s_m,w_m\right) $} and finally, for each pair $\left( t^n,w^n\right) $ we produce $2^{n\rho _2}$ independent $n$-sequences $v^n\left( t^n,w^n\right) \in \mathcal{V}^n$ according to {\small $\prod _{m=1}^n p\left( v_m |t_m,w_m\right) $}.\vspace{+.10cm}

\textit{Encoding at the transmitter:} For each pair $(s^n,t^n) \in T_{\delta}^{(n)}(S,T)$, we first find $\theta =\sigma (s^n)$, $\varphi =\tau (t^n)$ and $k^n=\alpha(s^n) =\beta (t^n)$. Then, we find a triplet {\small $\left( w^n_{\theta, \varphi}\left(k^n\right), u^n\left( s^n,w_{\theta, \varphi}^n\right) ,v\left( t^n,w_{\theta ,\varphi }^n\right) \right) $} so that {\small $\left( s^n, t^n, k^n, w^n_{\theta ,\varphi }\left( k^n\right) , u^n\left( s^n,w_{\theta, \varphi}^n\right), v^n\left(t^n,w_{\theta, \varphi}^n\right) \right) \in T_{\delta}^{(n)} \left(S,T,K,W,U,V\right)$}. Finally, we produce a $n$-sequence $x^n\left( s^n,t^n\right) \in \mathcal{X}^n$ according to $\prod _{m=1}^n p\left( x_m|w_m,u_m,v_m\right) $ for sending over the channel as the channel codeword of $\left( s^n,t^n\right) $. It is important to mention that if there is no such triplet {\small $\left( w^n_{\theta ,\varphi }\left( k^n\right) , u^n\left( s^n,w_{\theta ,\varphi }^n\right) ,v\left( t^n,w_{\theta ,\varphi }^n\right) \right) $} that {\small $\left( s^n, t^n, k^n, w^n_{\theta ,\varphi }\left( k^n\right) , u^n\left( s^n,w_{\theta ,\varphi }^n\right) ,v^n\left( t^n,w_{\theta ,\varphi }^n\right) \right) \in T_{\delta} ^{(n)} \left( S,T,K,W,U,V\right) $}, then an encoding error happens. However, according to lemma 2 in \cite{Han:BC}, if the following constraints are satisfied, then the existence of such a triplet for any $\delta >0$ and a given pair $\left( s^n,t^n\right) \in T_{\delta }^{(n)} \left( S,T\right) $ is guaranteed when $n$ is sufficiently large.
{\small 
\begin{align}
\label{1}
\rho _0 &> I\left( ST;W|K\right),\\
\label{2}
\rho _1 &> I\left( T;U|SW\right),  \\
\label{3}
\rho _2 &> I\left( S;V|TW\right), \\
\label{4}
\rho _1+\rho _2 &> I\left( SU;TV|W\right) -I\left( S;T|W\right) .
\end{align}}\vspace{-.40cm}

\textit{Encoding at the receivers:} As we stated before, we here use a one--round cooperation scheme. We consider $s_i\in [1:2^{nR_{s_i}}]$ as side information generated by receiver $i$, $i\in \{1,2\} $, to help the other receiver in the detection. We first consider the second receiver as a relay for the first receiver and describe the encoding. Each index $s_2$ is related to a codeword $x_2^n\left( s_2\right) $ which is produced $i.i.d.$ according to $\prod _{m=1}^n p\left( x_{2m}\right) $. For each $x_2^n\left( s_2\right) $, we produce $2^{nR_2}$ $i.i.d.$ $n$-sequences $\hat{y}^n_2$ according to 
{\small \begin{equation}
p\left( \hat{y}_2^n|x^n_2\left( s_2\right) \right) =\prod _{m=1}^np\left( \hat{y}_{2m}|x_{2m}\left( s_2\right) \right),\nonumber
\end{equation}
}%
and label these $n$-sequences $\hat{y}_2^n\left( z_2|s_2\right) $, $z_2\in [1:2^{n R_{2} }]$ and $s_2 \in \big [1:2^{nR_{s_2}}\big]$. Note that 
{\small \begin{align}
p\left( \hat{y}_2|x_2 \right) &=\sum_{\mathcal{X},\mathcal{Y}_{11},\mathcal{Y}_{21},\mathcal{Y}_{12}}p(x)p(y_{11},y_{21}|x)p(y_{12}|x_2)p(\hat{y}_2|y_{21},x_2),
\nonumber \\
p(x) &=\sum_{\mathcal{W},\mathcal{U},\mathcal{V}}p(w,u,v,x).\nonumber
\end{align}
}We partition the message set $\{1,\cdots,2^{nR_2}\}$ randomly and uniformly into $2^{nR_{s_2}}$ cells $S_{2\left( s_2\right) }$, $s_2 \in \big [1:2^{nR_{s_2}}\big]$. Assume that we want to transmit source information at block $i$. We use sequence $y_{21}^n$ received by the receiver 2 at block $(i-1)$ and find a triplet {\small $\Big( \hat{y}_2^n\left( z_{2}\left( i-1\right) |s_{2} \left( i-1\right) \right) , y_{21}^n\left( i-1\right), x_2^n\left( s_2\left( i-1\right) \right) \Big) \in T_{\delta }^{(n)}\left( X_2,Y_{21},\hat{Y}_2\right) $}, then we search for the cell $S_{2\left( s_{2}\left( i\right) \right) }$ for which $z_2(i-1) \in S_{2\left( s_{2}\left( i\right) \right)} $. Therefore, within the $i$th block of transmission (i.e., transmission of $\left( s^n\left( i\right) ,t^n\left( i\right) \right) $), the receiver 2 transmits the codeword related to index $s_2\left( i\right) $, i.e., $x_2^n\left( s_2\left( i\right) \right) $, to the receiver 1. Note that $s_2\left( i-1\right) $ is known from $z_2$ at block $\left( i-2\right) $.  
Similar encoding procedure is used by the receiver 1 with this difference that variables $S_{2\left( s_2\right) }$, $\hat{y}_2\left(z_2|s_2\right) $, $s_2\in \big [1:2^{nR_{s_2}}\big]$ and $z_2\in [1:2^{nR_2}]$ should be replaced by variables $S_{1\left( s_1\right) }$, $\hat{y}_1\left(z_1|s_1\right) $, $s_1\in \big [1:2^{nR_{s_1}}\big]$ and $z_1\in [1:2^{nR_1}]$, respectively.     

\subsection{Decoding scheme}
Recall that source information is to be sent at block $i$, and $y_1^n$ and $y_2^n$ (shown in Fig. \ref{Fig1}) are equal to $\left( y_{11},y_{12}\right)$ and $\left( y_{21},y_{22}\right)$, respectively.\vspace{+.10cm}

\textit{Decoding at the receivers:} Let us consider decoding procedure at receiver 1. Note that in this decoding procedure, information related to block $\left( i-1\right) $ is decoded at block $i$. Actually, detection at receiver 1 is done by having $y_1^n\left( i-1\right) $, $s_2\left( i\right) $ and $s_2\left( i-1\right) $. Decoding procedure at receiver 1 can be done in 3 steps as follows: 
\begin{enumerate}
\item Decoding of $s_2\left( i\right) $: Receiver 1 searchs for a $\hat{s}_2\left( i\right) $ such that $\left( x_2^n\left( \hat{s}_2\left( i\right) \right) ,y_{12}^n\left( i\right) \right) \in T_{\delta }^{(n)}$. According to the packing lemma, stated in \cite{ElGamal:book}, if we have the following inequality, then decoding of $s_2\left( i\right) $ will be done with small probability of error for $n$ large enough.
{\small \begin{equation}
R_{s_2}\leq I\left( X_2;Y_{12}\right).
\label{5}
\end{equation}
}%
Similarly, decoding of $s_1\left( i\right) $ at the receiver 2 is done with small probability of error for $n$ large enough if
{\small \begin{equation}
R_{s_1}\leq I\left( X_1;Y_{22}\right) .
\label{6}
\end{equation}}
\item Decoding of $z_2\left( i-1\right) $: First, we find a set $L_1\left( i-1\right) $ of $z_2$ related to block $\left( i-1\right) $ as below:
{\small \begin{IEEEeqnarray}{ll}
\Big\{ \hat{z}_2\in L_1\left( i-1\right) :& \Big( \hat{y}_2^n\left( \hat{z}_2|\hat{s}_2\left( i-1\right) \right) ,x_2^n\left( \hat{s}_2 \left( i-1\right) \right), \nonumber \\
& ~ y_{11}^n\left( i-1\right) ,y_{12}^n\left( i-1\right) \Big) \in T_{\delta }^{(n)} \Big\}, \nonumber 
\end{IEEEeqnarray}}%
next, we search for a $\hat{z}_2\left( i-1\right) $ such that
{\small \begin{equation}
\hat{z}_2\left( i-1\right) \in S_{2\left( \hat{s}_2\left( i\right) \right) }\bigcap L_1\left( i-1\right) , \nonumber
\end{equation}
}hence, decoded value of $z_2\left( i-1\right) $ is $\hat{z}_2\left( i-1\right) $. Similar procedure is done for decoding of $z_1\left( i-1\right) $ at receiver 2 by finding a set $L_2\left( i-1\right) $ of $z_1$, and $\hat{y}_2^n,\hat{s}_2,x_2^n,y_{11}^n$ and $y_{12}^n$ are replaced in above relations by $\hat{y}_1^n,\hat{s}_1,x_1^n,y_{21}^n$ and $y_{22}^n$, respectively.  
\item Detection of source information transmitted to receiver 1 i.e. $s^n\left( i-1\right) $: To this end, receiver 1 determines $\hat{s}^n\left( i-1\right) $ was sent, if it is the only element for which we have:
{\small \begin{IEEEeqnarray}{ll}
\big( s^n,k^n,w^n_{\theta ,\varphi}\left( k^n\right) ,u^n\left( s^n,w^n\right) , \nonumber \\
\hat{y}_2^n\left( \hat{z}_2\left( i-1\right) |\hat{s}_2\left( i-1\right) \right) ,y_{11}^n\left( i-1\right) ,y_{12}^n\left( i-1\right) \big) \in T_{\delta }^{(n)}. \nonumber
\end{IEEEeqnarray}}
Similar procedure is done for detection of $t^n\left( i-1\right) $ at receiver 2. Note that, to have estimated information at receiver 1 i.e. $\hat{s}^n\left( i-1\right) $ equal to real source information transmitted to this receiver i.e. $s^n\left( i-1\right) $, we must have $z_2\left( i-1\right) $ decoded correctly at receiver 1.
\end{enumerate} 

\textit{Decoding at the receivers act as relays:} First, let the receiver 2 be relay for the receiver 1. Within block $i$, receiver 2 (the relay) determines its cooperative message $\hat{z}_2\left( i\right) $ such that:
{\small \begin{equation}
\big( \hat{y}^n_2\left( \hat{z}_2\left( i\right) |\hat{s}_2\left( i\right) \right) ,y^n_{21}\left( i\right) ,x^n_2\left( \hat{s}_2\left( i\right) \right) \big) \in T_{\delta }^{(n)}, \nonumber
\end{equation}
}then, index $\hat{s}_2(i+1) $ for which $\hat{z}_2(i) \in S_{2\left( \hat{s}_2(i+1) \right)} $ is determined and its codeword $x^n_2\left( \hat{s}_2(i+1) \right) $ is prepared to be sent to receiver 2 within block $(i+1) $.    

Similar procedure is done for determination of $\hat{s}_1\left( i+1\right) $ at receiver 1 when it acts as a relay.
\subsection{Error events}
As stated earlier, within each block we decode source information related to previous block, so without loss of generality assume pair {\small $\left( s_0^n\left( i-1\right) ,t_0^n\left( i-1\right) \right) $} was sent within block {\small $\left( i-1\right) $} and we are in block $i$. We devide error events into two categories, events related to encoding and ones related to decoding.  

\textit{Encoding error events:} We have the following encoding error events:
\begin{enumerate}
\item If pair {\small $\left( s_0^n\left( i-1\right) ,t_0^n\left( i-1\right) \right) \in \mathcal{S}^n\times \mathcal{T}^n $}, i.e. output of sources, is not an element of {\small $T_{\delta }^{(n)}\left( S,T\right) $} then, encoding could not be done at the encoder. However, according to properties of strong typicality, probability of this event tends to zero when $n\rightarrow \infty $.

\item If for {\small $\left( s_0^n( i-1) ,t_0^n( i-1) \right) $} there is no triplet {\small $\left( w^n_{\left( \theta _0 ,{\varphi} _0\right) 0 }\left( k_0^n\right) , u_0^n\left( s_0^n,w_{\left( \theta _0,\varphi _0\right) 0}^n\right) ,v_0\left( t_0^n,w_{\left( \theta _0,\varphi _0\right) 0}^n\right) \right) $ }, in which $\theta _0=\sigma \left( s_0^n\right) $, $\varphi _0=\tau \left( t_0^n\right) $ and $k_0^n=\alpha \left( s_0^n\right) =\beta \left( t_0^n\right) $, such that:
{\small \begin{IEEEeqnarray}{ll}
\big( s_0^n, t_0^n, k_0^n, w^n_{\left( \theta _0,\varphi _0\right) 0}\left( k_0^n\right) , u_0^n\left( s_0^n,w_{\left( \theta _0,\varphi _0\right) 0}^n\right) , \nonumber \\
~~~ v_0^n\left( t_0^n,w_{\left( \theta _0,\varphi _0\right) 0}^n\right) \big) \in T_{\delta} ^{(n)} \left( S,T,K,W,U,V\right) ,\nonumber
\end{IEEEeqnarray}
}then, the encoder could not encode the information. According to relations $\left( \ref{1} \right) $-$\left( \ref{4}\right) $ for $\rho _0$, $\rho _1$ and $\rho _2$, probability of this error event is also arbitrarily small for large $n$. 
\end{enumerate}   

\textit{Decoding error events:} Assume no encoding error happens. We are at block $i$, actually we are going to analyze decoding of $s^n\left( i-1\right) $ and $t^n\left( i-1\right) $. We have following decoding error events:
\begin{enumerate}
\item Suppose that receiver $m$, $m\in \{1,2\}$, acts as relay. An error happens if cooperative message $z_m$ could not be found at this receiver, i.e. we have following event:
{\small \begin{IEEEeqnarray}{ll}
\{ \nexists z_m\in [1:2^{nR_m}]: \left( \hat{y}_m^n\left( z_m|\hat{s}_m\left( i-1\right) \right) ,y_{m1}^n\left( i-1\right) \right. , \nonumber \\
\left. x_m^n\left( \hat{s}_m( i-1) \right) \right) \in T_{\delta }^{(n)} \} ,\nonumber
\end{IEEEeqnarray}
}however, according to covering lemma, stated in details in \cite{ElGamal:book}, probability of this event will be arbitrarily small if we have:
{\small \begin{equation}
\label{7}
I\left( \hat{Y}_m;Y_{m1}|X_m\right) \leq R_m .
\end{equation}}
\item Assume above error event does not happen. Consider receiver $m$, $m\in\{1,2\}$, receives cooperative message from receiver $l$, $l\in \{ 2,1\}$, which plays the role of relay. If receiver $m$ could not estimate $s_l\left( i\right) $ correctly, then an error occurs, however as stated earlier when inequalities $\left( \ref{5}\right) $ and $\left( \ref{6}\right) $ hold, this estimation is done with small probability of error when $n$ is sufficiently large. Also, if receiver $m$ could not decode cooperative message correctly, i.e. we have following events, an error occurs:
{\small \begin{IEEEeqnarray}{ll}
E_{m1}=\{ z_l\left( i-1\right) \notin S_{l\left( \hat{s}_l\left( i\right) \right) }\bigcap L_1\left( i-1\right) \} \nonumber \\
E_{m2}=\{ \exists \tilde{z}_l\neq z_l\left( i-1\right) : 
\tilde{z}_l\in S_{l\left( \hat{s}_l\left( i\right) \right) }\bigcap L_1\left( i-1\right) \} .\nonumber
\end{IEEEeqnarray}
}Probability of event $E_{m1}$, according to Markov chain $Y_{m1},Y_{m2}\rightarrow Y_{l1}\rightarrow \hat{Y}_l ,X_l $ and Markov lemma, tends to zero when $n$ is sufficiently large. Also, according to packing lemma, probability of event $E_{m2}$ can be arbitrarily small if we have:
{\small \begin{equation}
\label{8}
R_l- R_{s_l} \leq I\left( \hat{Y}_l;Y_{m1},Y_{m2}|X_l\right) ,
\end{equation}
}when $n$ is sufficiently large.  
\item Assuming above events do not happen, we write error events may happen when we are decoding the information of sources at receivers. Consider source information decoding at receiver 1 which decodes sequences $s^n$. Receiver 1 declares $\hat{s}_0^n\left( i-1\right) $ as source information transmitted if there is only one element such that:
{\small \begin{IEEEeqnarray}{ll}
\left( \hat{s}^n_0( i-1) ,\hat{k}_0^n, w^n_{\left( \theta _0,\varphi _0\right) 0}\left( \hat{k}_0^n\right) , u_0^n\left( \hat{s}_0^n,w_{\left( \theta _0,\varphi _0\right) 0}^n\right) \right. ,\nonumber \\
 \hat{y}_2^n\left( \hat{z}_2\left( i-1\right) |\hat{s}_2\left( i-1\right) \right) ,y_{11}^n( i-1) ,y_{12}^n( i-1) \Big) \in T_{\delta }^{(n)}. \nonumber
\end{IEEEeqnarray}
}To write error events may happen here, first consider the following event:
{\small \begin{IEEEeqnarray}{lll}
E_{\theta ,\varphi ,r,p}\left( s^n\right) =\left( s^n\left( i-1\right) ,k^n, w^n_{\left( \theta ,\varphi \right) r}\left( k^n\right) , \right. ~~~~~~~\nonumber \\
~~~~~~~u_p^n\left( s^n,w_{\left( \theta ,\varphi \right) r}^n\right),\hat{y}_2^n\left( \hat{z}_2\left( i-1\right) |\hat{s}_2\left( i-1\right) \right) ,\nonumber \\
~~~~~~~ ~~~~~~~~~~~~~~y_{11}^n\left( i-1\right) ,y_{12}^n\left( i-1\right) \big) \in T_{\delta }^{(n)}. \nonumber
\end{IEEEeqnarray}
}We have the following error events considering {\small $\left( s^n_0\left( i-1\right) ,t_0^n\left( i-1\right) \right) $} is transmitted:
{\small
\begin{IEEEeqnarray}{lllll}
E_1^{(1)}= \{ E^c_{\theta _0,\varphi _0,0,0}\left( s_0^n\right)  \} \nonumber \\
E_2^{(1)}= \{ E_{\theta _0,\varphi _0,0,p}\left( s^n\right) \mathrm{for~some~} p \mathrm{~and~}s^n\mathrm{~with~} \nonumber \\
\sigma \left( s^n\right) =\theta _0,\alpha \left( s^n\right) =k^n_0, \left( p,s^n\right) \neq \left( 0,s_0^n\right) |E_{\theta _0,\varphi _0,0,0}\left( s_0^n\right) \} \nonumber \\
E_3^{(1)}=\{ E_{\theta ,\varphi ,r,p}\left( s^n\right) \mathrm{for~some~} \theta ,\varphi ,r,p\mathrm{~and~}s^n\mathrm{~with~} \nonumber \\
\theta =\sigma \left( s^n\right) ,\left( \theta ,\varphi ,r,\alpha \left( s^n\right) \right) \neq \left( \theta _0,\varphi _0,0,k_0^n \right) |E_{\theta _0,\varphi _0,0,0}\left( s_0^n\right) \} \nonumber
\end{IEEEeqnarray}
}where $c$ denotes complement. Similarly, for source information decoding at receiver 2, this receiver declares {\small $\hat{t}_0^n\left( i-1\right) $} was sent if there is only one element such that:
{\small \begin{IEEEeqnarray}{ll}
\left( \hat{t}^n_0\left( i-1\right) ,\hat{k}_0^n, w^n_{\left( \theta _0,\varphi _0\right) 0}\left( \hat{k}_0^n\right) , v_0^n\left( \hat{t}_0^n,w_{\left( \theta _0,\varphi _0\right) 0}^n\right) \right. ,\nonumber \\
\hat{y}_1^n\left( \hat{z}_1\left( i-1\right) |\hat{s}_1\left( i-1\right) \right) ,y_{21}^n\left( i-1\right) ,y_{22}^n\left( i-1\right) \Big) \in T_{\delta }^{(n)}. \nonumber
\end{IEEEeqnarray}
}Here, we consider the following event:
{\small \begin{IEEEeqnarray}{lll}
E_{\theta ,\varphi ,r,q}\left( t^n\right) =\left( t^n\left( i-1\right) ,k^n, w^n_{\left( \theta ,\varphi \right) r}\left( k^n\right) , \right. ~~~~~~~\nonumber \\
~~~~~~~~~~~~~~v_q^n\left( t^n,w_{\left( \theta ,\varphi \right) r}^n\right),\hat{y}_1^n\left( \hat{z}_1\left( i-1\right) |\hat{s}_1\left( i-1\right) \right) ,\nonumber \\
~~~~~~~~~~~~~~~~~~~ y_{21}^n\left( i-1\right) ,y_{22}^n\left( i-1\right) \big) \in T_{\delta }^{(n)}. \nonumber
\end{IEEEeqnarray} 
}As stated earlier, {\small $\left( s^n_0\left( i-1\right) ,t_0^n\left( i-1\right) \right) $} is transmitted, so we have:
{\small
\begin{IEEEeqnarray}{lllll}
E_1^{(2)}=\{ E^c_{\theta _0,\varphi _0,0,0}\left( t_0^n\right) \} \nonumber \\
E_2^{(2)}=\{ E_{\theta _0,\varphi _0,0,q}\left( t^n\right) \mathrm{for~some~} q \mathrm{~and~}t^n\mathrm{~with~} \nonumber \\
\tau \left( t^n\right) =\varphi _0,\beta \left( t^n\right) =k^n_0, \left( q,t^n\right) \neq \left( 0,t_0^n\right) |E_{\theta _0,\varphi _0,0,0}\left( t_0^n\right) \} \nonumber \\
E_3^{(2)}=Pr\{ E_{\theta ,\varphi ,r,q}\left( t^n\right) \mathrm{for~some~} \theta ,\varphi ,r,q\mathrm{~and~}t^n\mathrm{~with~} \nonumber \\
\varphi =\tau \left( t^n\right) ,\left( \theta ,\varphi ,r,\beta \left( t^n\right) \right) \neq \left( \theta _0,\varphi _0,0,k_0^n \right) |E_{\theta _0,\varphi _0,0,0}\left( t_0^n\right) \}. \nonumber
\end{IEEEeqnarray}
}According to properties of jointly strong typical sequences, probabilities of $E_1^{(1)}$ and $E_1^{(2)}$ could be arbitrarily small for $n$ large enough. For evaluation of probabilities $Pr\{ E_2^{(1)}\}$ and $Pr\{ E_2^{(2)}\}$ we do the same procedure as for $Q_1^{(2)}$ and $Q_2^{(2)}$ in [2-page 646], with $Y_1$ and $Y_2$ replaced with $\left( \hat{Y}_2,Y_{11},Y_{12}\right) $ and $\left( \hat{Y}_1,Y_{21},Y_{22}\right) $, respectively. So, $Pr\{ E_2^{(1)}\}$ tends to zero as $n\rightarrow \infty$ if we have:
{\small \begin{equation}
\label{9}
H\left( S|KW\right) \leq I\left( SU;\hat{Y}_2,Y_{11},Y_{12}|KW\right) +r_1-\rho _1 ,
\end{equation}
}and similarly $Pr\{ E_2^{(2)}\}$ tends to zero as $n\rightarrow \infty$ if we have:
{\small \begin{equation}
\label{10}
H\left( T|KW\right) \leq I\left( TV;\hat{Y}_1,Y_{21},Y_{22}|KW\right) +r_2-\rho _2 .
\end{equation}
}$Pr\{ E_3^{(1)}\}$ and $Pr\{ E_3^{(2)}\}$ are also evaluated as $Q_1^{(3)}$ and $Q_2^{(3)}$ in [2-page 646], respectively with $Y_1$ replaced with $\left( \hat{Y}_2,Y_{11},Y_{12}\right) $ and $Y_2$ replaced with $\left( \hat{Y}_1,Y_{21},Y_{22}\right) $. These error probabilities tend to zero as $n\rightarrow \infty $ if following inequalities hold, respectively:
{\small
\begin{IEEEeqnarray}{llll}
\label{11}
H\left( S|KW\right) +H\left( K\right) \leq I\left( SUW;\hat{Y}_2,Y_{11},Y_{12}\right)~~~~~~~~~~ \nonumber \\
~~~~~~~~~~~~~~~~~~~~~~~~~~~~~~~-r_2-\rho _0-\rho _1, \\
\label{12}
H\left( T|KW\right) +H\left( K\right) \leq I\left( TVW;\hat{Y}_1,Y_{21},Y_{22}\right)~~~~~~~~~~ \nonumber \\
~~~~~~~~~~~~~~~~~~~~~~~~~~~~~~~-r_1-\rho _0-\rho _2.
\end{IEEEeqnarray} } 
\end{enumerate}
Using Fourier-Moutzkin elimination algorithm and the fact that $0\leq r_1$ and $0\leq r_2$ and removing parameters $\rho _0, \rho _1,\rho _2, r_1, r_2, R_1$ and $R_2$, and considering inequalities $\left( \ref{1}\right) -\left( \ref{12}\right) $ relations in Theorem 1 are obtained.  


\section{Conclusion}
In this paper, we obtained a sufficient condition for arbitrarily correlated sources to be admissible for the DM-BC with one–-round cooperation between receivers through noisy channels. It was shown that our result subsumes some previous works as its special cases.

\end{document}